\documentclass[pmlr]{jmlr}

\usepackage{longtable}

\usepackage{booktabs}
\usepackage[load-configurations=version-1]{siunitx} 

\theorembodyfont{\upshape}
\theoremheaderfont{\scshape}
\theorempostheader{:}
\theoremsep{\newline}

\jmlrvolume{273}
\jmlryear{2025}
\jmlrworkshop{iRAISE25}

\title[M2LADS Demo]{M2LADS Demo: A System for Generating Multimodal Learning Analytics Dashboards}

\author{
  \Name{Alvaro Becerra}\textsuperscript{\normalfont 1} \Email{alvaro.becerra@uam.es} \vspace{-5mm}\\ 
  \AND
  \Name{Roberto Daza}\textsuperscript{\normalfont 2} \Email{roberto.daza@uam.es}  \vspace{-5mm}\\ 
  \AND
  \Name{Ruth Cobos}\textsuperscript{\normalfont 1} \Email{ruth.cobos@uam.es}  \vspace{-5mm}\\  
  \AND
  \Name{Aythami Morales}\textsuperscript{\normalfont 2} \Email{aythami.morales@uam.es}  \vspace{-5mm}\\ 
  \AND
  \Name{Julian Fierrez}\textsuperscript{\normalfont 2} \Email{julian.fierrez@uam.es} \\ 
  {\small
  \textsuperscript{\normalfont 1}{\normalfont \textit{GHIA, School of Engineering, Universidad Autónoma de Madrid}} \\
  \textsuperscript{\normalfont 2}{\normalfont \textit{BiDA-Lab, School of Engineering, Universidad Autónoma de Madrid}}
  }
}

\begin{document}

\maketitle

\begin{abstract}
We present a demonstration of a web-based system called M2LADS (``System for Generating Multimodal Learning Analytics Dashboards''), designed to integrate, synchronize, visualize, and analyze multimodal data recorded during computer-based learning sessions with biosensors. This system presents a range of biometric and behavioral data on web-based dashboards, providing detailed insights into various physiological and activity-based metrics. The multimodal data visualized include electroencephalogram (EEG) data for assessing attention and brain activity, heart rate metrics, eye-tracking data to measure visual attention, webcam video recordings, and activity logs of the monitored tasks. M2LADS aims to assist data scientists in two key ways: (1) by providing a comprehensive view of participants' experiences, displaying all data categorized by the activities in which participants are engaged, and (2) by synchronizing all biosignals and videos, facilitating easier data relabeling if any activity information contains errors.

\end{abstract}
\begin{keywords}
Biometrics and Behavior, Dashboard, Multimodal Learning Analytics, Online Learning, Web-based Technology
\end{keywords}

\section{Introduction}
\label{sec:intro}
The field of Multimodal Learning Analytics (MMLA) has expanded significantly in recent years, as the integration of physiological data with traditional behavioral metrics has created new opportunities to understand the dynamics of learning processes \citep{nagao2019artificial,  sharma2020multimodal, giannakos2022multimodal}.

New online learning platforms \citep{baro2018integration, hernandez2019edbb, nagao2023virtual, daza2025smartevr} have emerged that rely on biosensors and machine learning to better understand and predict learner behavior,offering insights to address challenges like limited learner-instructor interaction \citep{iraj2020understanding}, dropout rates \citep{topali2019exploring}, and distractions \citep{becerra2024biometrics}. However, due to the diversity of biosensors and the volume of biometric and behavioral signals that can be captured in monitoring sessions, accurate integration, synchronization, and visualization of multimodal data are essential. These processes ensure data integrity, address missing values, and verify correct data labeling. Furthermore, they provide an initial preview and analysis to identify patterns before applying more machine learning techniques. 
Our approach, M2LADS \citep{becerra2023m2lads, becerra2023user}, leverages web-based dashboards to provide a centralized and interactive system for processing, storing, visualizing, and analyzing multimodal data collected during the monitoring of learning sessions. By synchronizing signals with video recordings, M2LADS simplifies data relabeling and correction, providing a holistic view of learner engagement. The system integrates data from biosensors and biometric modules, such as attention level estimation \citep{daza2021alebk, daza2023matt, daza2024deepface}, heart rate \citep{hernandez2020heart}, eyeblink detection \citep{daza2020mebal, daza2024mebal2} and anomalous posture \citep{becerra2024biometrics}. This makes M2LADS a valuable tool for extracting meaningful insights from multimodal datasets \citep{navarro2024vaad, daza2024mebal2, daza2024improve}.


\section{M2LADS: System for Generating Multimodal Learning Analytics Dashboards}\label{sec:m2lads}

\begin{figure}[t]
    \centering
    \includegraphics[width=1\linewidth]{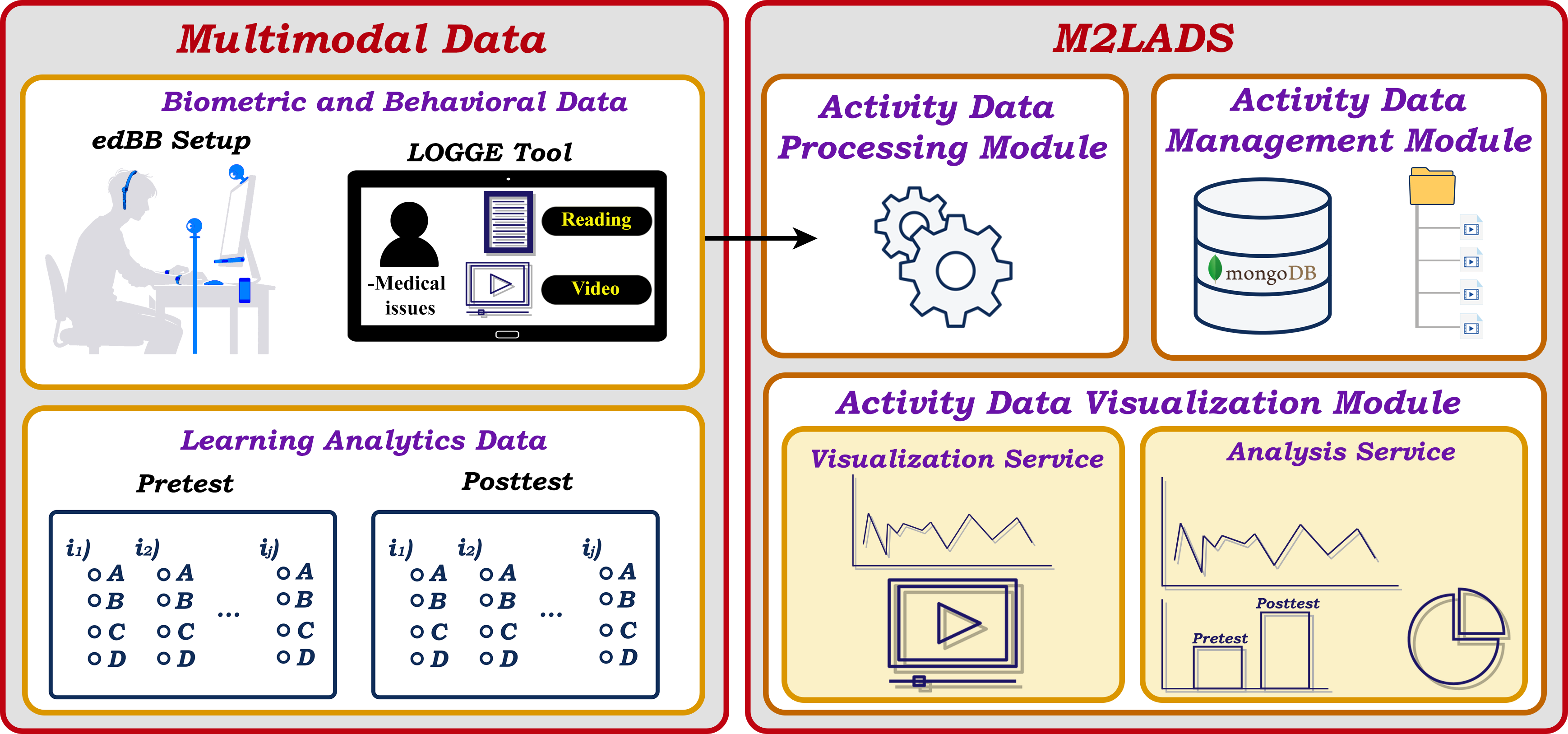}
    \caption{M2LADS arquitecture and modules}
    \label{fig:m2lads}
\end{figure}

Figure \ref{fig:m2lads} shows the three modules of the M2LADS system and the different multimodal data sources. M2LADS uses biometric data from two sources: 1) data captured with the edBB platform \citep{daza2023edbb}, which includes biosensors such as an EEG band for measuring attention, meditation, and brain waves; an eye tracker for visual attention videos and pupil diameter; two smartwatches for heart rate; and webcams in different positions capturing RGB and infrared images; and 2) data captured with the LOGGE tool \citep{becerra2023m2lads}, which includes information about the activities that the learners are engaged in during the monitoring sessions, along with demographic and medical data. Additionally, M2LADS can also use data from pretests and posttests that learners may take during a learning session.

\textbf{The Activity Data Processing Module} enables the extraction, cleaning, selection, and preprocessing of multimodal data to synchronize signals and videos and classify the values based on the activity the learner is engaged in. This module standardizes timestamps across all biosensors and data sources, ensuring temporal alignment. Furthermore, it classifies and map raw multimodal data to specific learning activities. For example, video data from webcams are segmented to match timestamps from LOGGE files, enabling precise identification of learner actions such as watching a video or solving an assignment. Similarly, biometric signals like attention levels, meditation states, and heart rate are linked to these activities, providing a holistic view of the learner's engagement and physiological response during each task. To enhance data analysis, the module also calculates smoothed data using a sliding window of 30 seconds. This technique helps to reduce noise and highlight trends in the biometric signals, making the data more interpretable and reliable for subsequent analysis. Additionally, this module calculates correlations between different data streams, enabling advanced insights that are highly valuable for machine learning models.

\textbf{The Activity Data Management Module} provides connectivity to MongoDB, where the processed biometric data are stored, along with directories containing audiovisual files. To ensure privacy and compliance with data protection regulations, all stored data are fully anonymized, with each student assigned a unique random identifier that replaces any personally identifiable information.

\textbf{The Activity Data Visualization Module} generates a personalized dashboard for each learner, providing a comprehensive representation of their activity data during the session. These dashboards are built using the Dash framework \footnote{\url{https://plotly.com/dash/}} and feature interactive visual components such as graphs. The visualizations include detailed graphs of the learner’s attention, meditation, heart rate, and neural wave patterns throughout the session, all categorized by the specific activity the learner was engaged in.

In addition to biometric data, the dashboards integrate multiple synchronized videos, such as screen recordings, webcam footage from various angles, and fixation area visualizations derived from the eye tracker. This synchronization ensures that instructors and researchers can easily cross-reference visual and biometric data, enhancing the interpretability of the learner's engagement and behavior.

The module also provides advanced analytical insights through specialized graphs, such as those that analyze which activities exhibit peaks and troughs in the signals, enabling the identification of activities that, for example, provoked higher levels of attention. It also includes correlation graphs that reveal relationships among biometric signals, as well as comparative analyses of pretest and posttest performance. These insights offer a deeper understanding of the learner’s progress and the impact of specific activities or interventions. For a demonstration of the dashboard’s capabilities, refer to the full video \footnote{See full video: \url{https://youtu.be/vwLAh-Gm4KU}}.

\section{Conclusion}


M2LADS has demonstrated its value as a versatile tool for researchers at our institution, supporting a wide range of studies on learner engagement and behavior during educational sessions. Notably, it has been utilized to validate biometric signals, compare learner performance metrics, and analyze signals based on activities, as highlighted in \cite{daza2024improve}. Furthermore, it has facilitated data relabeling efforts \citep{becerra2024biometrics,navarro2024vaad}, addressing the challenges posed by manually assigned labels during monitoring learning sessions.

\acks{Support by projects: HumanCAIC (TED2021-131787B-I00 MICINN), SNOLA (RED2022-134284-T), and BIO-PROCTORING (GNOSS Program, Agreement Ministerio de Defensa-UAM-FUAM dated 29-03-2022). A. Morales is also supported by the Madrid Government in the line of Excellence for University Teaching Staff (V PRICIT). Work conducted within the ELLIS Unit Madrid.}

\bibliography{pmlr_sample}

\end{document}